\documentclass[a4paper,11pt]{article}
\usepackage{pos}
\usepackage{gensymb}
\usepackage{overpic}

\usepackage[caption=false]{subfig}
\title{
	Muons as a tool for background rejection in imaging atmospheric Cherenkov telescope arrays}
 \ShortTitle{Background rejection with muons}

\author*[a]{L. Olivera-Nieto}
\author[b]{A. M. W. Mitchell}
\author[a]{K. Bernl\"{o}hr}
\author[a]{J. A. Hinton}

\affiliation[a]{Max Planck Institut f\"{u}r Kernphysik,\\
	Saupfercheckweg 1, 69117 Heidelberg, Germany}
\affiliation[b]{Department of Physics, ETH Zurich,\\
	CH-8093 Zurich, Switzerland}
\emailAdd{laura.olivera-nieto@mpi-hd.mpg.de}

\abstract{The presence of muons in air-showers initiated by cosmic ray protons and nuclei is well established as a powerful tool to separate such showers from those initiated by gamma-rays. However, so far this approach has been fully exploited only for ground level particle detecting arrays. In this contribution, we explore the feasibility of using Cherenkov light from muons as a background rejection tool for imaging atmospheric Cherenkov telescope arrays at the highest energies. We adopt an analytical model of the Cherenkov light from individual muons to allow rapid simulation of a large number of showers in a hybrid mode. This allows exploration of the very high background rejection power regime at acceptable cost in terms of computing time. We find that for very large telescopes ($\gtrsim$20 m diameter), efficient identification of muons would provide a major improvement with respect to standard background rejection techniques at energies above several tens of TeVs.}

\FullConference{37$^{\rm{th}}$ International Cosmic Ray Conference (ICRC 2021)\\
		July 12th -- 23rd, 2021\\
		Online -- Berlin, Germany}

\begin{document}
\maketitle

\section{Introduction}
Muons are produced in large numbers when charged cosmic ray particles travel through the atmosphere, primarily from the decay of charged pions. This fact has long been recognized as useful information to discriminate between hadronic and electromagnetic cascades, a critical distinction for gamma-ray astronomy~\cite{Gaisser}.

Muons become a relevant separation tool for showers above $\sim$1~TeV at high altitude~\cite{harmpaper}. However, excellent hadron rejection power, that is, over a factor $10^{4}$ reduction, is possible only at tens of~TeV. Detector arrays such as LHAASO~\cite{lhaaso-crab} have demonstrated the power of this approach through the deployment of extensive muon detector arrays, leading to a dramatic reduction in the background above tens and even hundreds of TeVs.

Imaging Atmospheric Cherenkov Telescope (IACT) arrays exploit the differences in shower structure between hadronic and electromagnetic showers to perform background rejection~\cite{DanStefan,stefanBDT,magicBDT}. This approach leads to superior rejection power at around $\sim$1~TeV, but performance worsens above few tens of TeVs~\cite{DanStefan,stefanBDT,dan-HE}. This is in part due to the fact that high-energy events are typically not fully contained by the camera, which hinders the calculation of the shower parameters used to discriminate the primary particle nature. This loss of performance at high energies can be seen, albeit indirectly, in the expected background rate after background rejection cuts for CTA\footnote{\url{https://www.cta-observatory.org/science/cta-performance/\#1472563453568-c1970a6e-2c0f}}.

Telescopes with a large mirror, such as the central telescope of the H.E.S.S. array~\cite{hess2} or the Large-Sized Telescopes (LSTs) of CTA~\cite{cta} are able to detect individual muons out to large impact distances due to their large collection area. This has been typically seen as a problem, as it is an additional source of background at low energies. This is because at large impact distances, muon images can resemble those produced by low-energy gamma rays. However, it could also be an opportunity to improve the background rejection power at the highest energies if muons are efficiently identified as such.

In this contribution we explore the potential for muon detection with IACTs as a tool for background rejection by characterizing the number of muons that are detectable by a large Cherenkov telescope in proton- and gamma-initiated showers of different energies.

\section{The simplified muon model (SMM)}
The Cherenkov emission of a single muon as it traverses the atmosphere, is straight-forward to characterize in comparison to that of air-showers in general. The suppressed bremsstrahlung cross-section of muons allows a majority of muons to reach ground-level with only ionisation losses. Similarly the reduced multiple scattering of muons means that the assumption of a linear trajectory is reasonable in most cases. This means that full simulation of muons might not be necessary to describe their Cherenokov emission, and a semi-analytical treatment is possible. This is important because, in order to explore the regime of muon-poor proton showers, very high statistics are needed, with the associated computing power becoming a limiting factor. 

The key parameters affecting the Cherenkov image properties of individual muons are the initial energy, production height in the atmosphere and impact distance to the telescope. Muons that hit the telescope dish produce ring-like images, with a smaller section of the ring forming an image as the distance of the muon away from the telescope dish becomes larger. The surface brightness of these images, however, remains mostly constant. This is used to calibrate Cherenkov detectors, and it also means that muons are able to trigger out to large distances, when the image no longer resembles a section of arc but rather a clump of pixels.

We assume a linear trajectory of the muon as it travels down from the production height, $h_{\mathrm{prod}}$. The path through the atmosphere is then $r = h/\cos\theta_z$, where $\theta_z$ is the muon zenith angle. We use atmospheric profiles for density $\rho(h)$ and refractive index $n(h)$ corresponding to the location of the H.E.S.S. array in Namibia~\cite{atmosphere}. We also take into account the wavelength-dependent atmospheric absorption A$(\lambda,h)$, integrated along the muon path. As the muon traverses the atmosphere with velocity $\beta(h)$, it produces Cherenkov photons once the Cherenkov condition ($\beta > \frac{1}{n(h)}$) is met. These photons, produced at a height $h$ arrive at a distance $R(h)$ from the muon impact point at ground
\begin{equation}
\label{eq:impact}
R(h) = (h-h_{\mathrm{ground}})\cdot\frac{\sin\theta_c}{\cos \theta_z}~,
\end{equation}
where $\theta_c = \arccos((n(h)\cdot \beta(h)^{-1})$ is the Cherenkov angle.
Next, we need to establish the number of photons collected by the telescope camera. This quantity is subject to the wavelength-dependent telescope response, referred to as W$_T(\lambda)$. Additionally to this, there are a number of wavelenght-independent geometric factors that combine to a factor $f\sim 0.6$ for a 28~m telescope.

Combining all the factors above with the well-known expression for the Cherenkov yield, we can compute the number of Cherenkov photons produced per unit path that are detected by the telescope:
\begin{equation}
\frac{dN_{tel}}{dr} = 2f\pi\alpha \left( 1-(\beta n)^{-2}\right)\int_{\lambda_1}^{\lambda_2} A(\lambda, h) W_T(\lambda)\frac{d\lambda}{\lambda^2}~,
\end{equation}
where $\alpha$ is the fine-structure constant.

\begin{figure*}
	\centering
    \begin{minipage}[c]{.5\textwidth}
	\centering
		\begin{overpic}[scale=0.38]{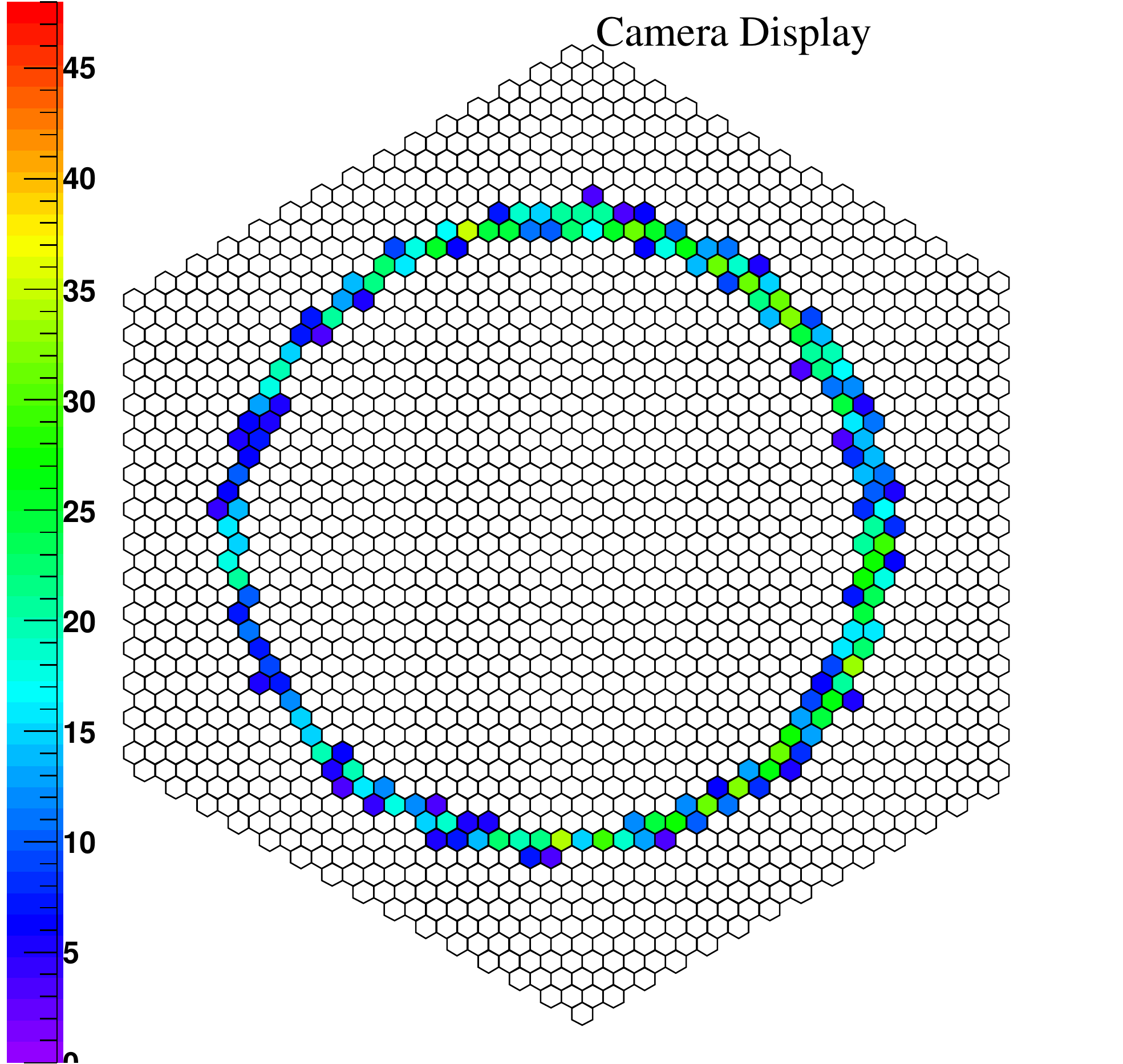}
			\put(85,85){R$\sim5$~m, E$\sim90$~GeV}
		\end{overpic}
\end{minipage}%
\begin{minipage}[c]{.5\textwidth}
	\centering
	\includegraphics[scale=0.38]{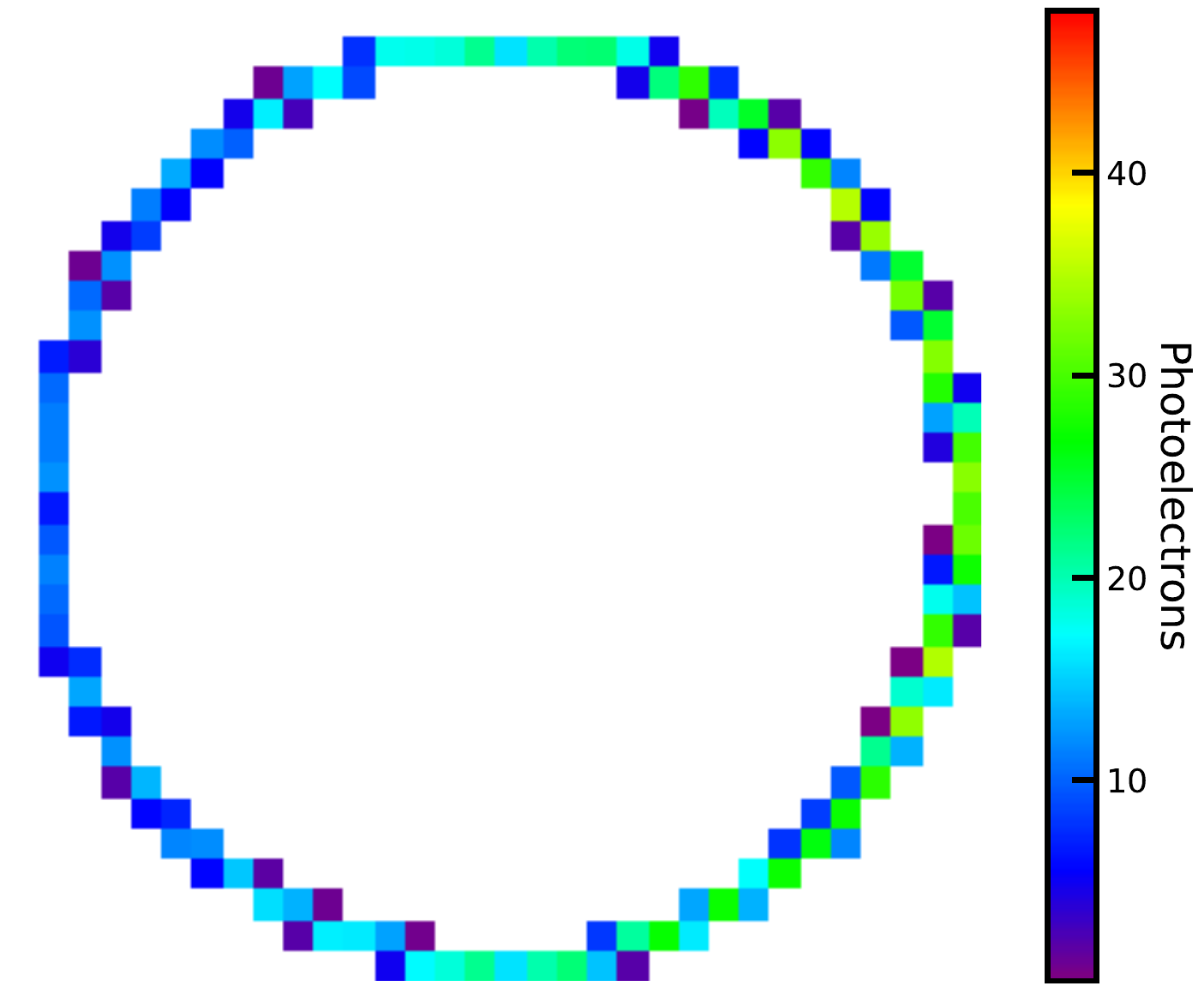} \\
\end{minipage}
    \begin{minipage}[c]{.5\textwidth}
	\centering

			\begin{overpic}[scale=0.38]{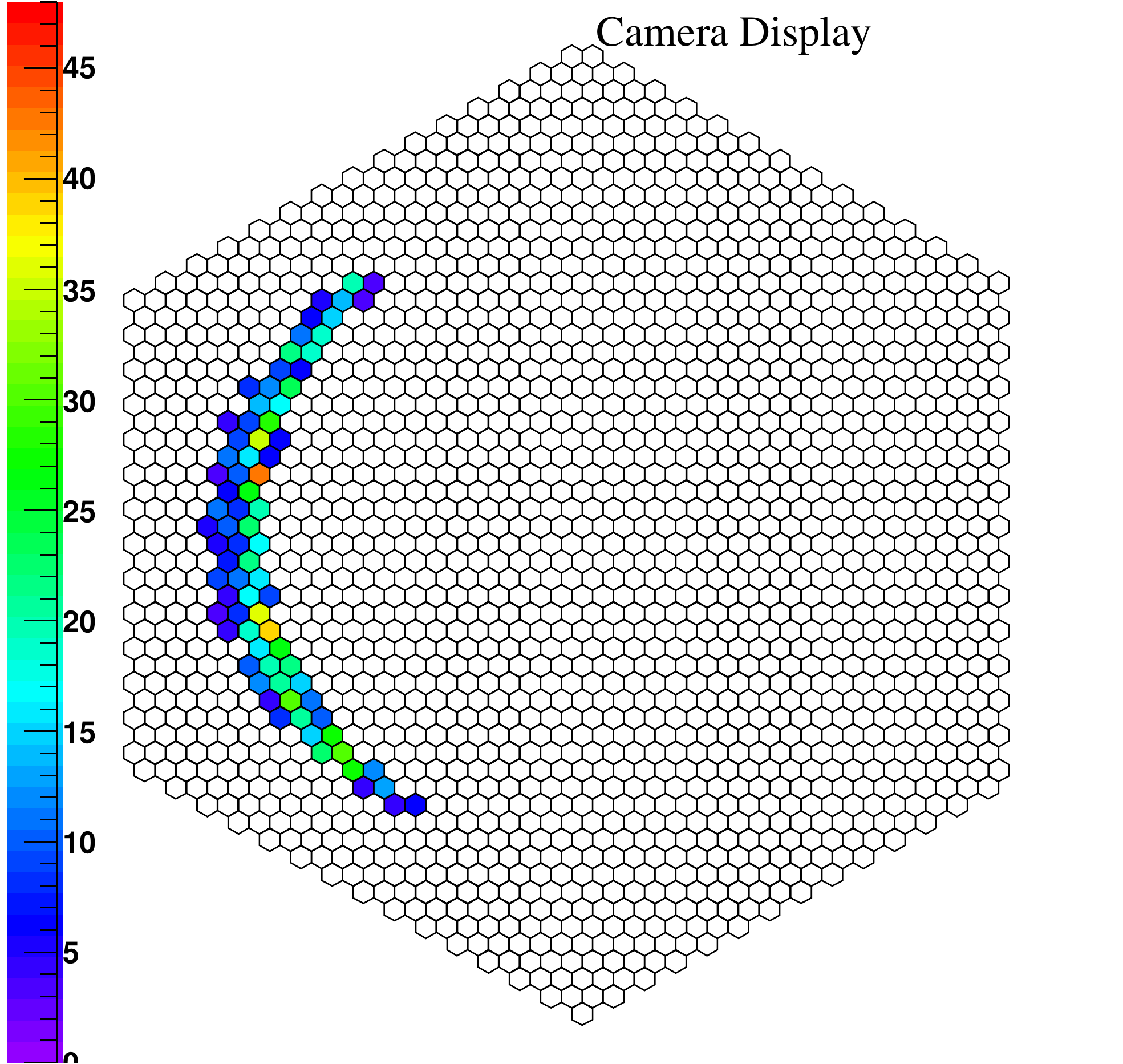}
		\put(85,85){R$\sim16$~m, E$\sim75$~GeV}
	\end{overpic}
\end{minipage}%
\begin{minipage}[c]{.5\textwidth}
	\centering
	\includegraphics[scale=0.38]{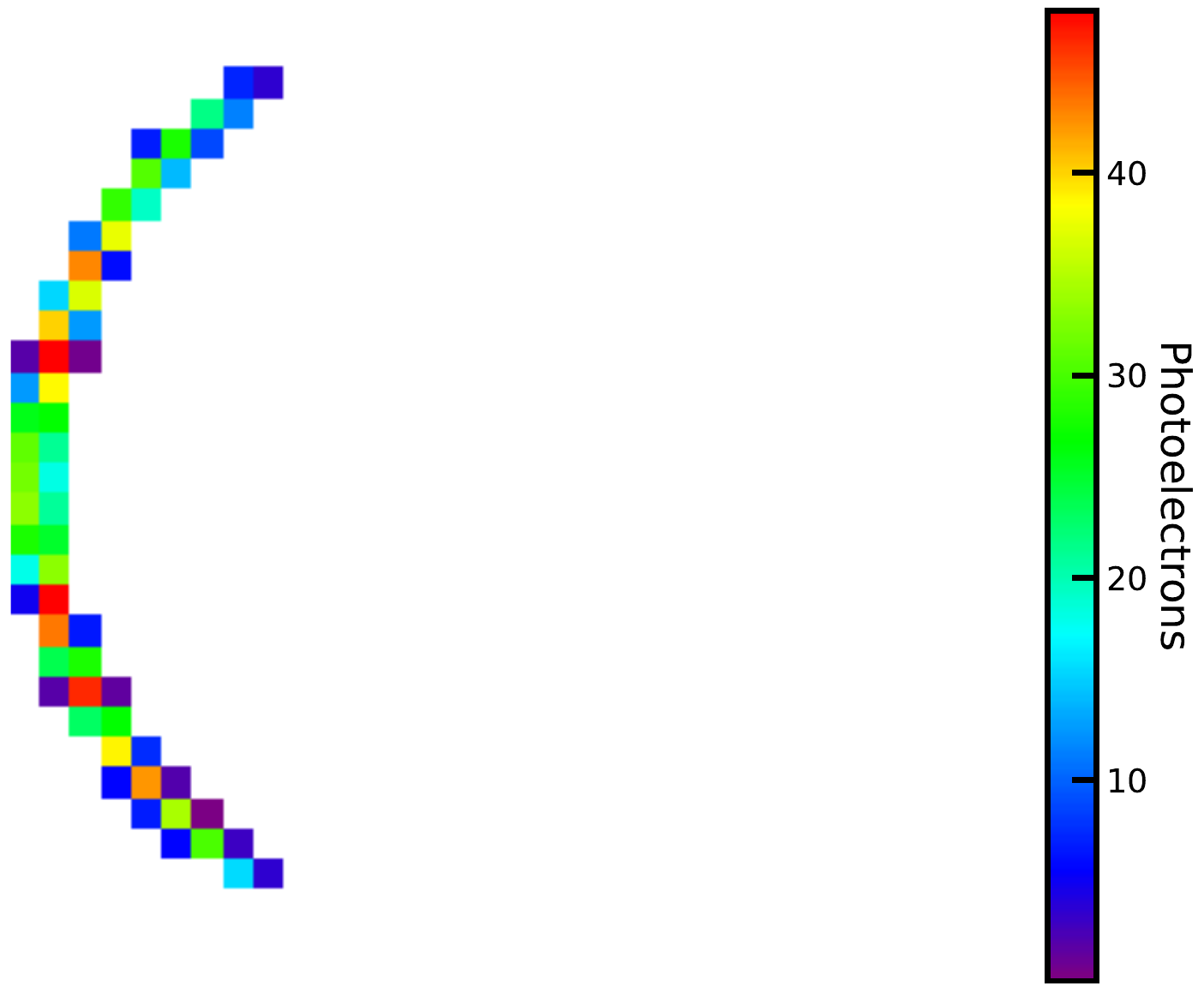} \\
\end{minipage}

	\caption{Comparison of muon images resulting from the full simulation of Cherenkov light via CORSIKA and \textit{sim\_telarray} (left) with the resulting images from the SMM (right) for different values of impact distance and energy. In both cases, the incoming muon zenith angle is 20\degree. The pixel shape is different, but the angular size is roughly of the same order. }
	\label{fig:muonimages}       % Give a unique label
\end{figure*}

We can integrate this distribution inside the telescope mirror to produce the Cherenkov image. We refer to this approach as the simplified muon model (SMM), from here on. The right column of Figure \ref{fig:muonimages} shows some examples of the resulting muon images. On the left column, for comparison, there are images for muons with the same properties produced with a full simulation chain, with CORSIKA~\cite{corsika} for the particle and Cherenkov light simulation and \textit{sim\_telarray}~\cite{sim-telarray} for the telescope simulation.

\begin{figure}[h]
 	\centering
   \includegraphics[width=0.65\textwidth]{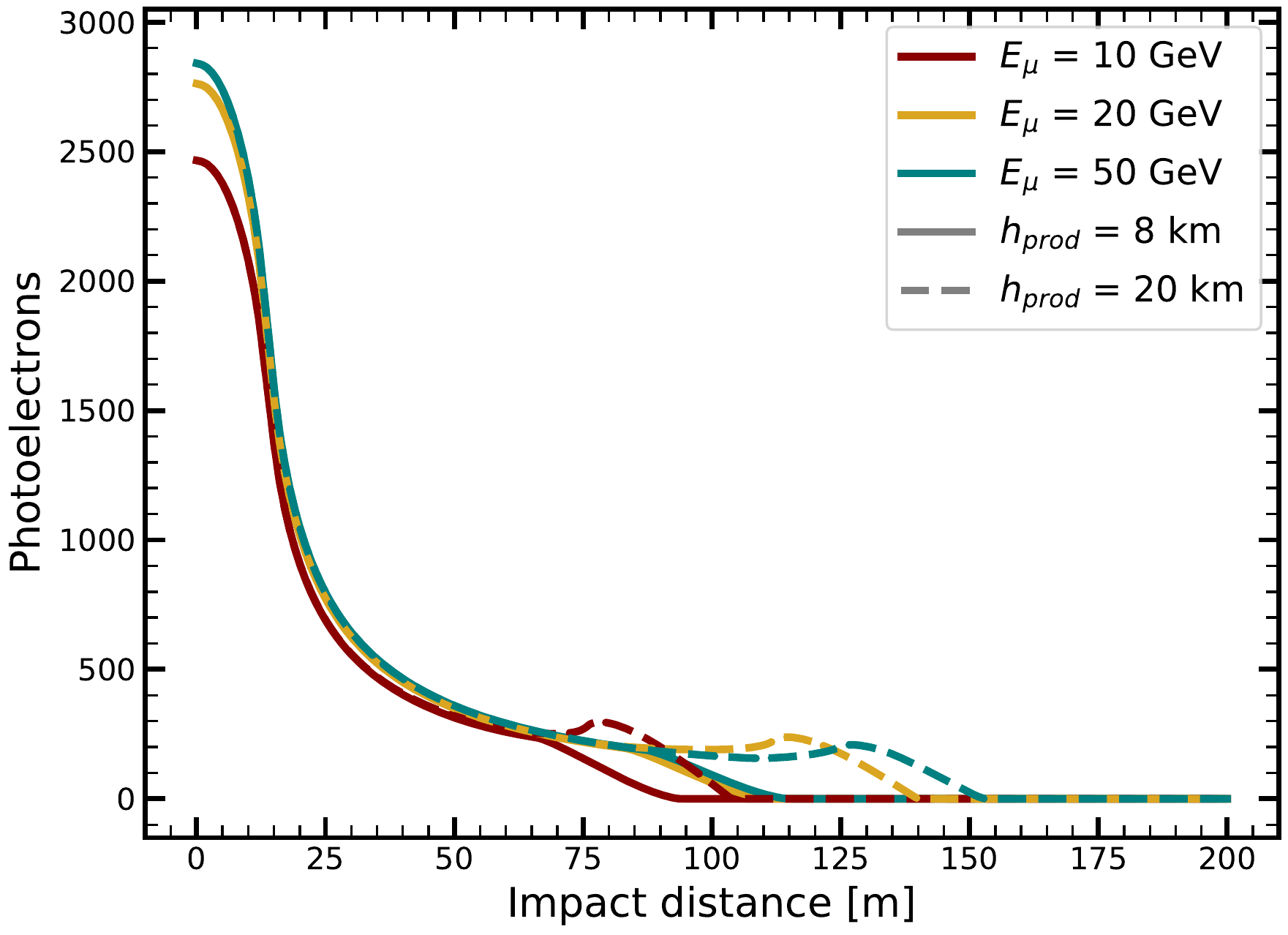}
 \caption{Amount of photoelectrons predicted by the SMM to be collected by a 28~m telescope for several muon energies and production heights.}
 \label{fig:ldfSMM}       % Give a unique label
 \end{figure}

From the images, we can compute the total number of photoelectrons collected by the camera as a function of impact distance for a muon with a given energy, incident angle and production height. This quantity will be used to determine whether a muon would be able to trigger the telescope camera. Figure~\ref{fig:ldfSMM} shows, for different values of initial muon energy and production height, the resulting number of photoelectrons collected as a function of ground impact distance.

\section{Detectable muons in showers}
We consider shower development only down to the Cherenkov threshold for muons, which allows for efficient simulations up to high primary energies with CORSIKA. For both gamma-ray- and proton-initiated showers, over $10^7$ showers were produced with energies between $10^1$ and $10^{2.5}$~TeV, distributed as $\propto E^{-1}$. A smaller sample of 3$\cdot 10^ 5$ iron-initiated showers was also produced. For all showers, the primary particle zenith angle was $\theta_z=20$\degree. For all the muons present in these showers, the production height, energy, ground level direction and impact point were extracted.

\begin{figure}[h]
	\centering
	\includegraphics[width=0.65\textwidth]{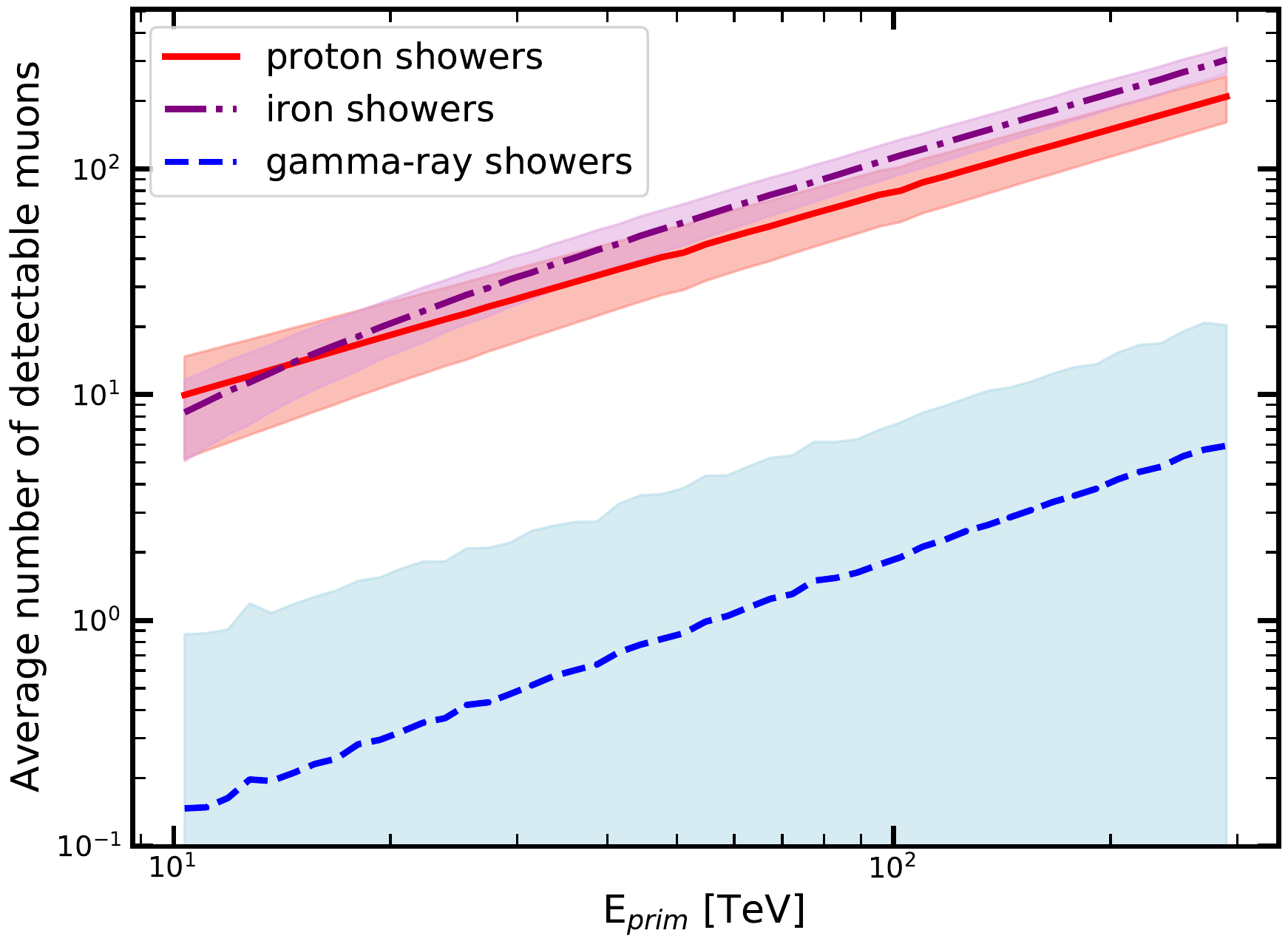}
	\caption{Average number of muons that pass all three detectability cuts for showers initiated by gamma-rays, protons and iron nuclei as a function of primary energy. Shaded areas represent the standard deviation.}
	\label{fig:result}       % Give a unique label
\end{figure}

We define a muon as \textit{detectable} if it passes three conditions:
\begin{enumerate}
	\item Enough photoelectrons reach the camera to activate the trigger. The trigger is the same as the one used by FlashCam, the camera installed in the largest H.E.S.S. telescope since October 2019~\cite{fc-trigger}. The criterion is passed when an image has a group of nine neighboring pixels with a total of more than 68 photoelectrons.
	\item The muon falls inside the telescope field of view (FoV). Assuming for simplicity that the telescope is always pointed towards the direction of the primary shower particle, this translates to a cut on the maximum angular distance between muon and primary particle of half the telescope FoV. The FoV is taken here to be that of the largest H.E.S.S. telescope, 3.5$\degree$.
	\item The muon is not too close to the main shower. This cut is applied to reflect the fact that while a muon can be detectable, it might be challenging to identify it as a muon. This is especially true if the images of the shower and muon overlap in the camera, although note that this is not necessarily an issue for timing-based identification methods. We apply a cut of 0.3$\degree$ as the minimum angular distance between the trajectories of the primary particle and the muon. Close to 50\% of muons that pass the other two cuts are removed by this criterion.
\end{enumerate}
From the initial list of muons per shower, we compute how many of them survive all three cuts, i.e., are \textit{detectable}. Figure~\ref{fig:result} shows the average number of muons that pass all three cuts as a function of primary energy for proton-, iron- gamma-ray-initiated showers. As can be seen, for all energies, the number of muons in hadronic showers that are, in principle, possible to detect with a large telescope is very high, particularly so at several tens of TeVs. On the other hand, muons produced in gamma-ray showers are comparably rare, which means that few gamma-ray showers contain more than one detectable muon.

\section{Conclusion}
Figure~\ref{fig:result} shows that significant potential exists for improving the background rejection performance of IACT arrays with at least one large telescope at high energies. This requires that individual muon arcs are identified efficiently in the presence of bright shower images. The result shown in Figure~\ref{fig:result} assumes a muon identification efficiency of around 50\%, which is probably optimistic. However, the large disparity in detectable muons between electromagnetic and hadronic showers means that even for more than an order of magnitude lower identification efficiency, there would be benefits to the muon-tagging approach.

The size of the telescope mirror is crucial to the detection of muons. Smaller telescopes have a lesser collection area and thus are unable to detect the faint light from muons out to large distances. The lower threshold of large Cherenkov telescopes is usually seen as an asset to the low-energy end of the spectrum. However, they can also potentially be critical to improve the high-energy performance of the array through such a muon-tagging approach, if included in the observations.

Showers produced by iron nuclei were included to take into account the fact that the hadronic background is in fact not solely made up by protons, but also heavy nuclei~\cite{crays}. The muon content in these showers is higher, which translates into a higher number of expected detectable muons as can be seen in Figure~\ref{fig:result}. In a sense, the proton case is the worst-case scenario, being the hadronic background species most likely to initiate muon-poor showers. However, protons are also the most likely species to masquerade as gamma ray events in traditional background rejection approaches~\cite{2007APh....28...72M}, and therefore the room for improvement is greater.

A forthcoming publication will describe this analysis in detail, including several checks of the SMM. This will include the full exploration of the number muons in showers, down to the very-rare muon-poor proton showers. We look forward to releasing these results to the community in due course.
\bibliographystyle{JHEP}
\bibliography{refs} 
%% Full authors list (ONLY FOR COLLABORATIONS)
%\clearpage
%\section*{Full Authors List: \Coll\ Collaboration}
%
%\noindent \textbf{Note comment afterwards:} Collaborations have the possibility to provide an authors list in xml format which will be used while generating the DOI entries making the full authors list searchable in databases like Inspire HEP. For instructions please go to icrc2021.desy.de/proceedings or contact us under icrc2021proc@desy.de.\\
%
%\scriptsize
%\noindent
%first.author$^1$, 
%second.author$^2$, 
%third.author$^3$ % .... more names
%and 
%last.author$^{n}$ \\
%
%\noindent
%$^1$first.affiliation.
%$^2$second.affiliation. % .... more affiliation
%$^{m}$last.affiliation.

\end{document}